\documentclass[english,aps,preprint]{revtex4}
\usepackage[T1]{fontenc}
\usepackage[latin9]{inputenc}
\usepackage{float}
\usepackage{bm}
\usepackage{amsmath}
\usepackage{amssymb}
\usepackage{graphicx}
\usepackage{esint}
\usepackage {ushort}
\usepackage{xcolor}
\usepackage{ulem}

\makeatletter

\@ifundefined{textcolor}{}
{%
 \definecolor{BLACK}{gray}{0}
 \definecolor{WHITE}{gray}{1}
 \definecolor{RED}{rgb}{1,0,0}
 \definecolor{GREEN}{rgb}{0,1,0}
 \definecolor{BLUE}{rgb}{0,0,1}
 \definecolor{CYAN}{cmyk}{1,0,0,0}
 \definecolor{MAGENTA}{cmyk}{0,1,0,0}
 \definecolor{YELLOW}{cmyk}{0,0,1,0}
}

\begin{document}


\vspace*{1cm}

\title{Ground state of many-electron systems based\\ on the action function}

\author{P. Fulde}
\email{fulde@pks.mpg.de}
\affiliation{Max-Planck-Institut f\"ur Physik komplexer Systeme, N\"othnitzer Stra\ss e 38, 01187 Dresden, Germany}
\date{\today}

\vspace*{2cm}

\begin{abstract}
The Hilbert space for an interacting electron system increases exponentially with electron number $N$. This limits the concept of wavefunctions $\psi$ based on solutions of the Schr\"odinger equation to $N \leq N_0$ with $N_0 \simeq 10^3$ \cite{Kohn1999}. It is  argued that this exponential wall problem (EWP) is connected with an increasing  redundance of information contained, e.g., in the ground-state of the system and it's wavefunction. The EWP as well as redundance of information are avoided when the characterization of the ground state is based on the action function $R$ rather than on the 
solutions $\psi$ of the Sch\"odinger equation. Both are related through a logarithm, i.e., $R = -i \hbar \ ln \psi$. Working with the logarithm is made possible by the use of cumulants. It is pointed out the way electronic structure calculations for periodic solids may use this concept.
\end{abstract}

\maketitle


It is well known that the wave character of nonrelativistic electrons is described by the Schr\"odinger equation and its solutions. It is also known that the dimension of the associated Hilbert space is increasing exponentially with the electron number $N$. This brings up the question whether this exponential increase limits the concept of wavefunctions based on the Schr\"odinger equation. Indeed, the following statement has been made by Walter Kohn in his Nobel lecture in 1999 \cite{Kohn1999}: for a system with $N > N_0$ electrons where typically $N \approx 10^3$ a wavefunction $\psi ({\bf r}_1 \sigma_1, \dots, {\bf r}_N \sigma_N)$ is no longer a legitimate scientific concept! This statement is well founded. The Schr\"odinger equation for a many-electron system cannot be  solved analytically. Therefore one must be able to rely one approximate solutions. Consider, e.g., the ground state $| \psi_0 \rangle$ of a many-electron system. It is noticed that the overlap with any approximate state $| \psi_0^{\rm app} \rangle$ vanishes exponentially with $N$, i.e.,
\begin{equation*}
\mid \left< \psi_0 | \psi_0^{\rm app}\right> \mid \leq \left( 1 - \epsilon \right)^N 
\label{eq:01}
\end{equation*}
for large $N$. Here $\epsilon$ is an acceptable inaccuracy in the description of an electron. Thus for $N > N_0$ any chosen state is orthogonal to the exact one, which however cannot be determined. A similar argument applies to another important requirement, i.e., that we must be able to document the state $| \psi_0 \rangle$. We would need an exponentially increasing number of entries in order to specify the parameters entering it. 

The exponential wall problem (EWP), i.e., the limitation of wavefunctions to $N \leq  N_0$ does not show up when the electronic interactions are either 
neglected or treated in a mean-field approximation, as in a Hartree-Fock \cite{Hartree1928,Evarestov06} or Kohn-Sham \cite{Kohn1999}, or N\'eel state. Here the ground state can be described by a single configuration, e.g., a Slater determinant. Tacitly, a closed shell nondegenerate state is assumed here.

The limitation of Schr\"odinger wavefunctions to $N < N_0$ is clearly unsatisfactory. Therefore formulations of the theory are desirable which invalidate this restriction. In the following we point out the way this is done.

An analysis of the origin of the EWP leads soon to the insight that it is the multiplicative character of the wavefunction when two nearly uncoupled subsystems A and B are considered which is responsible for the limitation, i.e., $| \psi (A, B) \rangle = | \psi (A) \rangle \otimes | \psi (B) \rangle$. Coupled with it is an increasing redundance of information contained in $| \psi (A,  B) \rangle$ when $N$ increases. For illustration consider $L$ atoms, e.g.,  He atoms with a total of $N = 2L$ electrons. The ground-state $| \psi_0 \rangle$ is of the general form 
\begin{equation}
| \psi_0 \rangle = \sum_{i_1, \dots, i_L} C_{i_1 \dots i_L} | i_1 \rangle \otimes | i_2 \rangle \otimes \dots \otimes | i_L \rangle
\label{eq:02}
\end{equation}
where $| i_\nu \rangle$ labels the different configurations of electrons on site $\nu$. When the atoms are nearly uncoupled the matrix in front factorizes in
\begin{equation}
\label{eq:03}
C_{i_1 \dots i_L} = c_{i_1} \cdot c_{i_2} \dots c_{i_L}~~~.
\end{equation}
Let us denote by $m$ the number of configurations with which we want to describe the two electrons in the ground state of a single He atom. The total number of configurations is according to (\ref{eq:03}) given by $m^L$ and therefore exponentially increasing with $L = N/2$. Yet, the information contained in (\ref{eq:03}) is simply that contained in a single set $c_{i \nu}$, supplemented with the information that all atoms are equal. Therefore one may argue that it is the redundance of information which is the origin of the EWP. A theory which wants to avoid the EWP must therefore also get rid of this redundance. This is the case when a wavefunction is additive instead of multiplicative when the two subsystems $A$ and $B$ are considered. This is achieved by defining wavefunctions with the help of action functions. They are related to the solutions of a Schr\"odinger equation through a logarithm. This is outlined in the following. Before this, we recall briefly the derivation of Schr\"odinger's equation.

Already Hamilton realized that the action function of a classical particle, i.e., $W({\bf r}, t) = -Et + R ({\bf r})$ with $E$ denoting the particle energy, behaves like the phase does in geometrical optics. Stated differently, the equation for action waves is of the same form as the eikonal equation in optics. Therefore Schr\"odinger started for the derivation of an equation for matter waves from the ansatz
\begin{equation}
\label{eq:04}
\Psi = e^{iW/\hbar} = e^{i(-Et + R)/ \hbar}~~.
\end{equation}
By setting
\begin{equation}
\label{eq:10}
R = -i \hbar \ ln  \psi
\end{equation}
and assuming for $\Psi$ a wave equation he obtained for $\psi$ the Schr\"odinger equation. The energy $E$ as well of the action function $R$ are additive with respect to the subsystems $A$ and $B$ and therefore $\psi$ is multiplicative. Thus when $R/\hbar$ is used instead of $\psi$ for the definition of the ground state and it's wavefunction, the latter does not face an EWP. Neither is there any redundance of information contained in it.

Working with a logarithm is avoided by making use of cumulants \cite{Kubo1962,Fulde1995}. An example is a classical nonideal gas. The additive free energy $F$ is related to the multiplicative partition function $Z$ through $F \sim ln Z$. The contribution of a pair interaction $U$ to $F$ is obtained by a cumulant expansion (Mayer's cluster expansion \cite{Mayer40}). The general of the cumulant of a quantum mechanical matrix element is
\begin{equation}
\label{eq:05}
\left< \phi_1 \left| A_1 \dots A_M \right| \phi_2 \right>^c = \frac{\partial}{\partial_{\lambda_1}} \dots \frac{\partial}{\partial_{\lambda_M}} ln \left. \left< \phi_1 \left| \prod^M_{i = 1} e^{\lambda_i A_i} \right| \phi_2 \right> \right|_{\lambda_i = 0} ~~~,
\end{equation}
where it has been assumed that $\langle \phi_1 | \phi_2 \rangle \neq 0$. The $A_L$ denote arbitrary operators. In the simplest case it is seen that
\begin{equation*}
\label{eq:09}
ln \left< \phi_1 \left| e^{\lambda A} \right| \phi_2 \right> = \left< \phi_1 \left| e^{\lambda A} -1 \right| \phi_2 \right>^c~~~.
\end{equation*}
The characteristic feature of cumulants is the suppression of statistically independent contributions to the matrix element. Thus all operator contractions must be {\it connected} when the matrix element is evaluated.

In order to take advantage of cumulants for the definition of the ground state of a many-electron system based on the action function we proceed as follows.We split the Hamiltonian $H$ into two parts $H = H_0 + H_1$. The part $H_0$ is an effective one-particle operator, e.g., a Hartree-Fock or Kohn-Sham or N\'eel operator. The corresponding ground state $| \Phi_0 \rangle$ is assumed to be known, i.e., $H_0 | \Phi_0 \rangle = \epsilon_0 | \Phi_0 \rangle$. In the following we shall call $| \Phi_0 \rangle$ the vacuum state. The remaining part $H_1$ of $H$ generates vacuum {\it fluctuations}. They are represented by operators acting on $| \Phi_0 \rangle$ and are elements of an operator space. A well known example is the configurational interaction representation of the ground state $| \psi_0 \rangle$ of $H$
\begin{equation}
\label{eq:06}
| \psi_0 \rangle = \left( 1 + \sum_{i \mu} \alpha^i_\mu c^+_i c_\mu + \sum_{i<j \atop \mu< \nu} \alpha^{ij}_{\mu \nu} c^+_i c^+_j c_\nu c_\mu + \dots \right) | \Phi_0 \rangle = \tilde{\Omega} | \Phi_0 \rangle~~,
\end{equation}
Here $c^+_i (c_\mu)$ are creation (annihilation) operators of electrons in spin orbitals $i(\mu)$. The operator $\tilde \Omega$ is the wave- or M\o ller operator which transforms the ground state of $H_0$ to the exact one.

We want to express the ground state of a many-electron system in terms of  vacuum fluctuations based on the action function (\ref{eq:10}). We denote this state by $| \psi_0 \rangle^c$. In distinction to $| \psi_0 \rangle$ we require that any matrix element formed with $| \psi_0 \rangle^c$ is {\it additive} when separated subsystems are considered. This implies that these matrix elements do not contain factorizable contributions. This is achieved by setting
\begin{eqnarray}
\label{eq:07}
| \psi_0 \rangle ^c & = & | \Omega \Phi_0 \rangle^c \nonumber\\
& = & | \Omega )~~.
\end{eqnarray}
and requiring that in operator space the following bilinear form (i.e., metric) is used
\begin{eqnarray*}
\label{eq:11}
(A|B) = \left< \Phi_0 \left| A^+ B \right| \Phi_0 \right>^c~~~.
\end{eqnarray*}
Here $A$ and $B$ are arbitrary operators. Equation (\ref{eq:07}) expresses the ground state in operator space.

From \cite{Fulde19} it is seen that $| \Omega ) = | 1 + S)$. The {\it connected} vacuum fluctuations are described by $| S)$. There is no longer redundant information contained in $| S)$ because of its additive properties. The EWP does not exist for $| \Omega )$ \cite{Fulde1995,Fulde19}. Note that the correlation energy is $E_{\rm corr} = (H|S)$. Similar, for any operator $A$ it is $\langle \psi_0 | A | \psi_0  \rangle / \langle \psi_0 | \psi_0 \rangle = (\Omega | A \Omega)$.

\begin{figure}
\includegraphics[width=0.62\textwidth]{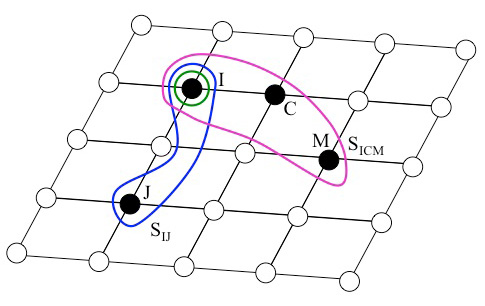}
\caption{Examples of connected vacuum fluctuations as described by different contributions $S_I, S_{IJ}$ and $S_{\rm ICM}$ to the cumulant scattering matrx  $| S )$ (see (\ref{eq:07}))
\label{fig:01}}
\end{figure}

This brings us to the documentation of $| S)$. For the ground state of a periodic solid the connected vacuum fluctuations can be calculated stepwise. We decompose $| S)$ in
\begin{equation}
\label{eq:09}
| S ) =  \sum_I | S_I ) + \sum_{\langle IJ \rangle} \left. \left| S_{IJ} - S_I - S_J \right. \right) + \dots~~~,
\end{equation}
where $I$, $J$ are site indices of the lattice \cite{Fulde1995}. In each case only a small number of electrons has to be considered, namely those situated in form of Wannier spin orbitals at sites $I$, $I$ and $J$ etc. The latter are particularly well localized for solids with a large gap in the excitation spectrum, i.e., for semiconductors and insulators. Metals with partially filled bands require additional measures \cite{Paulus11}. Because of the limited number of electrons involved, one may apply standard quantum-chemical methods such as Coupled Cluster or Coupled Electron Pair Approximations \cite{Evarestov06,Paulus11} to compute the different contributions to $| S)$. This is schematically shown in Illustration \ref{fig:01}. Numerical calculations have shown that connected vacuum fluctuations involving more than two sites $I$ and $J$ are rapidly decreasing. A rapid decrease is also found when the distance between two sites $I$ and $J$ is increased. Finally, we want to point out that strong correlations of electrons, e.g., on site $I$ can be modeled by vacuum fluctuations $| S_I )$ so that they correspond to a CASSCF (Complete Active Space Self-consistent Field) calculation.

Until here the ground state of a many-body electron system has been considered and it was shown how the EWP can be avoided. Similar considerations can also be applied to excited states.

\vspace{2cm}

I would like to thank K. Becker, H. Stoll and P. Thalmeier for numerous fruitful discussions.

\end{document}